\documentclass[conference,10pt,draftclsnofoot,onecolumn]{IEEEtran}

\usepackage{benstyle_IEEE}

\usepackage{algorithmic}
\usepackage{array}
\usepackage{url}

\begin{document}


\title{Optimal Sparse Recovery for Multi-Sensor Measurements}

\author{\IEEEauthorblockN{Il Yong Chun}
\IEEEauthorblockA{Department of Mathematics\\ Purdue University\\
West Lafayette, IN 47907 USA\\
Email: chuni@purdue.edu}
\and
\IEEEauthorblockN{Ben Adcock}
\IEEEauthorblockA{Department of Mathematics\\ Simon Fraser University\\
Burnaby, BC V5A 1S6 Canada\\
Email: ben\_adcock@sfu.ca}}


\maketitle

\thispagestyle{plain}
\pagestyle{plain}

\begin{abstract}
Many practical sensing applications involve multiple sensors simultaneously acquiring measurements of a single object.  Conversely, most existing sparse recovery guarantees in compressed sensing concern only single-sensor acquisition scenarios.  In this paper, we address the optimal recovery of compressible signals from multi-sensor measurements using compressed sensing techniques, thereby confirming the benefits of multi- over single-sensor environments.  Throughout the paper we consider a broad class of sensing matrices, and two fundamentally different sampling scenarios (distinct and identical respectively), both of which are relevant to applications.  
For the case of diagonal sensor profile matrices (which characterize environmental conditions between a source and the sensors), this paper presents two key improvements over existing results.  First, a simpler optimal recovery guarantee for distinct sampling, and second, an improved recovery guarantee for identical sampling, based on the so-called sparsity in levels signal model.
\end{abstract}


%
\IEEEpeerreviewmaketitle

\section{Introduction}
The standard single-sensor problem in compressed sensing (CS) involves the recovery of a sparse signal $x \in \bbC^{N}$ from measurements 
\be{
\label{standard_CS}
y = A x +e,
}
where $A \in \bbC^{m \times N}$ and $e \in \bbC^{m}$ is noise.  As is well known, subject to appropriate conditions on the measurement matrix $A$ (e.g.\ incoherence) it is possible to recover $x$ from a number of measurements that scales linearly with its sparsity $s$.

\subsection{System models}
In this paper, we consider the generalization of \R{standard_CS} to a so-called \textit{parallel acquisition} system \cite{Chun&Adcock:16arXiv-CS&PA}, where $C \geq 1$ sensors simultaneously measure $x$:
\be{
\label{parallel_CS}
y = A x + e,\qquad A = \left [ \begin{array}{c} A_1 \\ \vdots \\ A_C \end{array} \right ],\qquad y =  \left [ \begin{array}{c} y_1 \\ \vdots \\ y_C \end{array} \right ],\qquad e =  \left [ \begin{array}{c} e_1 \\ \vdots \\ e_C \end{array} \right ].
}
Here $A_{c} \in \bbC^{m_c \times N}$ is the matrix corresponding to the measurements taken in the $c^{\rth}$ sensor and $e_c \in \bbC^{m_c}$ is noise.  Throughout, we assume that
\bes{
A_c = \tilde{A}_{c} H_c,
}  
where $\tilde{A}_{c} \in \bbC^{m_c \times N}$ are standard CS matrices (e.g.\ a random subgaussian, subsampled isometry or random convolution), and $H_c \in \bbC^{N \times N}$ are fixed, deterministic matrices, referred to as \textit{sensor profile} matrices.  These matrices model environmental conditions in the sensing problem; for example, a communication channel between $x$ and the sensors, the geometric position of the sensors relative to $x$, or the effectiveness of the sensors to $x$.    As in standard CS, our recovery algorithm will be basis pursuit:
\be{
\label{recovery_alg}
\min_{z \in \bbC^N} \| z \|_{1}\ \mbox{subject to $\| A z - y \|_2 \leq \eta$},
}
Here $\eta > 0$ is such that $\| e \|_2 \leq \eta$.

Within this setup we consider two distinct types of problem:
\begin{itemize}
\item \textit{Distinct sampling.} Here the matrices $\tilde{A}_1,\ldots,\tilde{A}_C$ are independent; that is, drawn independently from possibly different distributions.
\item \textit{Identical sampling.} Here $m_1 = \ldots = m_C = m/C$ and $\tilde{A}_{1} = \ldots = \tilde{A}_C = \tilde{A} \in \bbC^{m/C \times N}$, where $\tilde{A}$ is a standard CS matrix.  That is, the measurement process in each sensor is identical, the only difference being in the sensor profiles $H_c$.
\end{itemize}

\subsection{Applications} \label{sec:Appl}

Parallel acquisition systems are found in numerous applications, and are employed for a variety of different reasons.

\subsubsection{Parallel magnetic resonance imaging} \label{sec:appl:pMRI}
Parallel MRI (pMRI) techniques are commonly used over single-coil MRI to reduce scan duration.  The most general system model in pMRI is an example of identical sampling with diagonal sensor profiles \cite{Chun&Adcock&Talavage:15TMI, Chun&Adcock&Talavage:14EMBS_pMRI, Pruessmann&Weiger&Scheidegger&Scheidegger:99MRM}.  In this case, the model \R{parallel_CS}--\R{recovery_alg} is the well-known CS SENSE technique for pMRI \cite{Chun&Adcock&Talavage:14EMBS_pMRI, Chun&Adcock&Talavage:15TMI, Knoll&Clason&Bredies&Uecker&Stollberger:12MRM, She&Chen&Liang&DiBella&Ying:14MRM}.

\subsubsection{Multi-view imaging}
In multi-view imaging -- with applications to satellite imaging, remote sensing, super-resolution imaging and more -- $C$ cameras with differing alignments simultaneously image a single object.  Following the work of of \cite{Park&Wakin:12EJASP, Traonmilin&etal:15JMIV}, this can be understood in terms of the above framework, with the sensor profiles $H_c$ corresponding to geometric features of the scene.

\subsubsection{Generalized sampling}
Papoulis' generalized sampling theorem \cite{Papoulis:77TCS,Unser:PIEEE} is a well-known extension of the classical Shannon Sampling theorem in which a bandlimited signal is recovered from samples of $C$ convolutions of the original signal taken at a lower rate (precisely $1/C$ of the Nyquist rate).  Common examples include jittered or derivative sampling, with applications to super-resolution and seismic imaging respectively.  Our identical sampling framework gives rises to a sparse, discrete version of generalized sampling.

\subsubsection{Other applications}
Besides acquisition time or cost reduction (e.g.\ pMRI and generalized sampling) or the recovery of higher-dimensional/resolution signals (e.g.\ in multi-view or light-field imaging), parallel acquisition systems are also used for power reduction (e.g.\ in wireless sensor networks), and also naturally arise in a number of other applications, including system identification.  We refer to \cite{Chun&Adcock:16arXiv-CS&PA} for details.

\subsection{Contributions}
The work \cite{Chun&Adcock:16arXiv-CS&PA} introduced the first CS framework and theoretical analysis for the system \R{parallel_CS}--\R{recovery_alg}.  We refer to this paper for further information and background.  Our work builds on this paper by introducing new recovery guarantees for the identical and distinct sampling scenarios.  Specifically, in Corollaries \ref{c:distinct_bound} and \ref{c:identical_bound} respectively we present new sufficient conditions on the sensor profile matrices $H_c$ so that the total number of required measurements $m$ is linear in the sparsity $s$ and independent of the number of sensors $C$.  Since this implies that the average number of measurements required per sensor $m_{\mathrm{avg}} = (m_1+\ldots+m_C)/C$ behaves like $s/C$, these results provide a theoretical foundation for the successful use of CS in the aforementioned applications.  To verify our recovery guarantees we provide numerical results showing phase transition curves.

\subsection{Notation}
Write $\nm{\cdot}_{p}$ for the $\ell^p$-norm on $\bbC^{N}$ and denote the canonical basis by $\{ e_i \}^{N}_{i=1}$.  If $\Delta \subseteq \{1,\ldots,N\}$ then we use the notation $P_{\Delta}$ for both the orthogonal projection $P_{\Delta} \in \bbC^{N \times N}$ with
\bes{
(P_{\Delta}x)_{j} = \left \{ \begin{array}{cc} x_j & j \in \Delta \\ 0 & \mbox{otherwise} \end{array} \right .,\qquad x \in \bbC^N,
}
and the matrix $P_{\Delta} \in \bbC^{|\Delta| \times N}$ with
\bes{
(P_{\Delta}x)_{j} = x_j,\quad j \in \Delta,\qquad x \in \bbC^N.
}
The conversion of a vector into a diagonal matrix is denoted by $\mathrm{diag}(\cdot)$.
Distinct from the index $i$, we denote the imaginary unit by $\I$.  
In addition, we use the notation $A \lesssim B$ or $A \gtrsim B$ to mean there exists a constant $c>0$ independent of all relevant parameters (in particular, the number of sensors $C$) such that $A \leq c B$ or $A \geq c B$ respectively.

\section{Abstract framework}
Following \cite{Chun&Adcock:16arXiv-CS&PA}, we now introduce an abstract framework that is sufficiently general to include both the identical and distinct sampling scenarios.  For more details we refer to \cite{Chun&Adcock:16arXiv-CS&PA}.

\subsection{Setup}\label{sec:general_setup}

For some $M \in \bbN$, let $F$ be a distribution of $N \times M$ complex matrices.  We assume that $F$ is isotropic in the sense that
\be{
\label{F_iso}
\bbE (B B^*) = I,\qquad B \sim F.
}
If $p = m/M$ (assumed to be an integer), let $B_1,\ldots,B_p$ be a sequence of i.i.d.\ random matrices drawn from $F$.  Then we define the measurement matrix $A$ by
\be{
\label{A_F}
A = \frac{1}{\sqrt{p}} \sum^{p}_{i=1} e_i \otimes B^*_i = \frac{1}{\sqrt{p}} \left [ \begin{array}{c} B^*_1 \\ \vdots \\ B^*_p \end{array} \right ] \in \bbC^{m \times N},
}
where $\otimes$ denotes the Kronecker product.

This framework is an extension of the well-known setup of \cite{Candes&Plan:11IT} for standard single-sensor CS, which corresponds to isotropic distributions of complex vectors (i.e.\ $M=1$), to arbitrary $N \times M$ matrices.  It is sufficiently general to allow us to consider both the distinct and identical sampling scenarios: 

\subsubsection{Distinct sampling, $M=1$}
\label{sec:setupDist}
In the $c^{\rth}$ sensor, suppose that the sampling arises from random draws from an isotropic distribution $G_c$ on $\bbC^N$.  Define $F_c$ so that $a_c \sim F_c$ if $a_c = H^*_c \tilde{a}_c$ for $\tilde{a}_c \sim G_c$.  Now let $X$ be a uniformly-distributed random variable taking values in $\{1,\ldots,C\}$.  Then define the distribution $F$ on $\bbC^N$ so that, when conditioned on the event $\{ X = c \}$, we have $F = F_c$.  Since $F$ should be isotropic in the sense of \R{F_iso}, this means that we require the \textit{joint isometry condition} $C^{-1} \sum^{C}_{c=1} H^*_c H_c = I$ for the sensor profiles.

\subsubsection{Identical sampling, $M=C$}
\label{sec:setupIdent}
Let $G$ be an isotropic distribution of vectors on $\bbC^N$.  Define the distribution $F$ on $\bbC^{N \times C}$ so that $B \sim F$ if $B = [H^*_1 a | \cdots | H^*_C a ]$ for $a \sim G$.
In this case, we require the joint isometry condition $\sum^{C}_{c=1} H^*_c H_c = I$ to satisfy the condition \R{F_iso}.

\subsection{Signal model}
As discussed in \cite{Chun&Adcock:16arXiv-CS&PA}, is it often not possible in multi-sensor systems to recover all sparse vectors with an optimal measurement condition.  This is due for the potential of clustering of the nonzeros of a sparse vector.  Instead, we shall consider a signal model that prohibits such clustering:

\defn{[Sparsity in levels]
\label{d:sparsity_lev}
Let $\cI = \{ I_1,\ldots,I_D \}$ be a partition of $\{1,\ldots,N\}$ and $\cS = (s_1,\ldots,s_D) \in \bbN^D$ where $s_d \leq | I_d|$, $d=1,\ldots,D$.  A vector $z \in \bbC^N$ is $(\cS,\cI)$-sparse in levels if $\left | \left \{ j : z_j \neq 0 \right \} \cap I_d \right | \leq s_d$ for $d=1,\ldots,D$.
}

Note that sparsity in levels was first introduced in \cite{Adcock&Hansen&Poon&Pi&Roman:13arXiv} as a way to consider the asymptotic sparsity of wavelet coefficients (see also \cite{Adcock&Hansen&Roman:14arXiv1403.6541}).

\defn{[Sparse and distributed vectors]
\label{d:sparse_distrib}
Let $\cI = \{ I_1,\ldots,I_D \}$ be a partition of $\{1,\ldots,N\}$ and $1 \leq s \leq N$.  For $1 \leq \lambda \leq D$, we say that an $s$-sparse vector $z \in \bbC^N$ is sparse and $\lambda$-distributed with respect to the levels $\cI$ if $z$ is $(\cS,\cI)$-sparse in levels for some $\cS = (s_1,\ldots,s_D)$ satisfying
\bes{
\max_{d=1,\ldots,D} \{ s_d \} \leq \lambda s / D.
}
We denote the set of such vectors as $\Sigma_{s,\lambda,\cI}$ and, for an arbitrary $x \in \bbC^N$, write $\sigma_{s,\lambda,\cI}(x)_1$ for the $\ell_1$-norm error of the best approximation of $x$ by a vector in $\Sigma_{s,\lambda,\cI}$.
}
Note that our interest lies with the case where $\lambda$ is independent of $D$; that is, when the none of the local sparsities $s_d$ greatly exceeds the average $s/D$.

\subsection{Abstract recovery guarantee}
Our first result concerns the recovery of the an arbitrary support set $\Delta$.  For this, we require the following (see \cite{Chun&Adcock:16arXiv-CS&PA}):

\defn{[Coherence relative to $\Delta$ ]
\label{d:coh_rel}
Let $F$ be as in \S \ref{sec:general_setup}  and $\Delta \subseteq \{ 1,\ldots,N\}$.  The local coherence of $F$ relative to $\Delta$ is $\Gamma(F,\Delta) = \max \left \{ \Gamma_{1}(F,\Delta),\Gamma_{2}(F,\Delta) \right \}$, where $\Gamma_i(F,\Delta)$, $i=1,2$, are the smallest numbers such that 
\bes{
\| B B^* P_{\Delta} \|_{\infty} \leq \Gamma_1(F,\Delta),\qquad B \sim F,
}
and
\bes{
\sup_{\substack{z \in \bbC^N \\ \| z \|_{\infty} = 1}} \max_{i=1,\ldots,N} \bbE | e^*_i B B^* P_{\Delta} z |^2 \leq \Gamma_{2}(F,\Delta),\qquad B \sim F,
}
almost surely.
}

\thm{[Abstract recovery theorem \cite{Chun&Adcock:16arXiv-CS&PA}]
\label{t:abs_recov}
For $N,M,p \in \bbN$ with $N \geq 2$ and $p M \leq N$ let $F$ be a distribution on $\bbC^{N \times M}$ satisfying \R{F_iso} and suppose that $0 < \epsilon < 1$, $\eta \geq 0$ and $\Delta \subseteq \{1,\ldots,N\}$ with $s = |\Delta| \geq 2$.  Let $x \in \bbC^N$ and draw $A \in \bbC^{m \times N}$ according to \R{A_F}, where $m=pD$.  Then for any minimizer $\hat{x}$ of
\bes{
\min_{z \in \bbC^N} \| z \|_{1}\ \mbox{subject to $\| A z - y \|_2 \leq \eta$},
}
where $y = A x + e$ with $\| e \|_2 \leq \eta$, we have
\bes{
\label{stab_robust_recovery}
\| x - \hat{x} \|_{2} \lesssim \| x - P_{\Delta} x \|_{1} + \sqrt{s} \eta,
}
with probability at least $1-\epsilon$, provided
\bes{
m \gtrsim M \cdot \Gamma(F,\Delta) \cdot L,
}
where
\bes{
\label{L_def}
L =  \log(N/\epsilon) + \log(s) \log(s/\epsilon) .
}
}

\section{Results for diagonal sensor profile matrices}

\subsection{Distinct sampling}
Let $H_c = \mathrm{diag}(h_c)$, $h_c = \{ h_{c,i} \}^{N}_{i=1} \in \bbC^N$, be diagonal sensor profiles.  We now let $M=1$ and $G_c$, $F_{c}$ be as in \S \ref{sec:setupDist}.  For simplicity, we suppose that $m_c = m/C$ for $c=1,\ldots,C$. We also assume that the distributions $G_1,\ldots,G_C$ are incoherent, i.e.\ $\mu(G_{c}) \lesssim 1$ for $c=1,\ldots,C$.\footnote{The \textit{coherence} $\mu(F)$ of a distribution $F$ of vectors in $\bbC^N$ is defined as the smallest number such that $\| a \|^2_{\infty} \leq \mu(F)$ almost surely for $a \sim F$ \cite{Candes&Plan:11IT}.}

\cor{
\label{c:distinct_bound}
Let $\cI = \{ I_1,\ldots,I_D\}$ be a partition of $\{1,\ldots,N\}$, $1 \leq \lambda \leq D$, and $2 \leq s \leq N$.  Let $x \in \bbC^{N}$, $0 < \epsilon < 1$ and $H_{c} \in \bbC^{N \times N}$, $c=1,\ldots,C$, be diagonal matrices satisfying the joint isometry condition
\bes{
\label{channel_iso:dist}
C^{-1} \sum^{C}_{c=1} H^*_c H_c = I.
}
Let $F$ be as in \S \ref{sec:setupDist} and draw $A$ according to \R{A_F}.  If $y = A x + e$, $\| e \|_{2} \leq \eta$, then for any minimizer $\hat{x}$ of
\bes{
\min_{z \in \bbC^N} \| z \|_{1}\ \mbox{subject to $\| A z - y \|_2 \leq \eta$},
}
we have
\bes{
\| x - \hat{x} \|_{2} \lesssim \sigma_{s,\lambda,\cI}(x)_1 + \sqrt{s} \eta,
}
with probability at least $1-\epsilon$, provided
\bes{
m \gtrsim \lambda \cdot s \cdot \mu \cdot \Upsilon_{\mathrm{distinct}} \cdot L,
}
where $\mu = \max_{c=1,\ldots,C} \mu(G_c)$ and
\bes{
\Upsilon_{\mathrm{distinct}} =  D^{-1} \max_{c=1,\ldots,C}  \sum^{D}_{d=1} \| h_c \|_{\infty} \| P_{I_d} h_c \|_{\infty}.
}
}

\prf{
By Theorem \ref{t:abs_recov} it suffices to estimate the coherence $\Gamma(F,\Delta)$ for subsets $\Delta$ of the form $\Delta = \Delta_1 \cup \cdots \cup \Delta_D$, where $\Delta_d \subseteq I_d$ and $s_d = | \Delta_d | \leq \lambda s / D$ for $d=1,\ldots,D$.  Fix $z \in \bbC^N$, $\| z \|_{\infty} =1$ and $i \in \{1,\ldots,N\}$.  If $B = H^*_c \tilde{a}_c$, where $\tilde{a}_c \sim G_c$ then
\eas{
|e^*_i B B^* P_{\Delta} z | &= \left | \sum^{D}_{d=1} \sum_{j \in \Delta_d } e^*_i H^*_c \tilde{a}_c \tilde{a}^*_{c} H_c e_j z_j \right |
\\
& \leq \| \tilde{a}_c \|^{2}_{\infty} \sum^{D}_{d=1} \sum_{j \in \Delta_d} | h_{c,i} | | h_{c,j} | z_j |
\\
& \leq \mu(G_c) \sum^{D}_{d=1} s_d \| h_{c} \|_{\infty} \| P_{I_d} h_c \|_{\infty}.
}
Hence $\Gamma_{1}(F,\Delta) \leq \lambda \cdot s \cdot \mu \cdot \Upsilon_{\mathrm{distinct}}$.  Also,
\eas{
\bbE |e^*_i B B^* P_{\Delta} z |^2 &= C^{-1} \sum^{C}_{c=1} \bbE \left |  e^*_i H^*_c \tilde{a}_c \tilde{a}^*_{c} H_c P_{\Delta} z \right |^2
\\
& \leq\mu  C^{-1} \sum^{C}_{c=1} |h_{c,i} |^2 \bbE | \tilde{a}^*_c H_c P_{\Delta} z |^2
\\
& \leq \mu  \left ( C^{-1} \sum^{C}_{c=1} |h_{c,i} |^2 \right ) \max_{c=1,\ldots,C} \| H_c P_{\Delta} z \|^2,
}
where in the last step we use the fact that $G_c$ is isotropic.  Observe that
\eas{
\| H_c P_{\Delta} z \|^2 &= \sum^{D}_{d=1} \sum_{j \in \Delta_d} | h_{c,j} |^2 |z_j|^2
\\
& \leq \sum^{D}_{d=1} s_d \| P_{I_d} h_c \|^2_{\infty}
\\
& \leq \frac{\lambda s}{D} \sum^{D}_{d=1} \| h_c \|_{\infty}  \| P_{I_d} h_c \|_{\infty}.
}
Also, the normalization condition $C^{-1} \sum^{C}_{c=1} H^*_c H_c = I$ implies that $C^{-1} \sum^{C}_{c=1} |h_{c,i} |^2 = 1$.  Substituting these into the previous bound now gives $\Gamma_{2}(F,\Delta) \leq \lambda \cdot s \cdot \mu \cdot \Upsilon_{\mathrm{distinct}}$.
To complete the proof, we now let $\Delta_{d}$ be the index set of the largest $s_d$ entries of $x$ restricted to $I_d$, where $s_d$ satisfies $s_d \leq \lambda s / D$.  Then $\| x - P_{\Delta} x \|_{1} = \sigma_{s,\lambda,\cI}(x)_1$ as required.
}

We remark that Corollary \ref{c:distinct_bound} is simpler than our previous result \cite[Cor.\ 3.5]{Chun&Adcock:16arXiv-CS&PA}.  Specifically, it requires only one condition on the sensor profile matrices.

\subsection{Identical sampling}
Now let $M=C$, $p=m/C$ and $G$, $F$ be as in \S \ref{sec:general_setup}.  We shall assume that $G$ is incoherent; $\mu(G) \lesssim 1$.

\cor{
\label{c:identical_bound}
Let $\cI = \{ I_1,\ldots,I_D\}$ be a partition of $\{1,\ldots,N\}$, $1 \leq \lambda \leq D$, and $2 \leq s \leq N$.  Let $x \in \bbC^{N}$, $0 < \epsilon < 1$ and $H_{c} \in \bbC^{N \times N}$, $c=1,\ldots,C$, be diagonal matrices satisfying the joint isometry condition
\bes{
\label{channel_iso:ident}
\sum^{C}_{c=1} H^*_c H_c = I.
}
Let $F$ be defined as in \S \ref{sec:setupIdent} and draw $A$ according to \R{A_F}.  If $y = A x + e$, $\| e \|_{2} \leq \eta$, then for any minimizer $\hat{x}$ of
\bes{
\min_{z \in \bbC^N} \| z \|_{1}\ \mbox{subject to $\| A z - y \|_2 \leq \eta$},
}
we have
\bes{
\| x - \hat{x} \|_{2} \lesssim \sigma_{s,\lambda,\cI}(x)_1 + \sqrt{s} \eta,
}
with probability at least $1-\epsilon$, provided
\bes{
m \gtrsim \lambda \cdot s \cdot \mu \cdot \Upsilon_{\mathrm{identical}} \cdot L,
}
where $\mu = \mu(G)$ and
\bes{
\Upsilon_{\mathrm{identical}} = \frac{C}{D} \max_{i=1,\ldots,N} \sum^{D}_{d=1} \max_{j \in I_d} \left | \sum^{C}_{c=1} \overline{h_{c,i}} h_{c,j} \right |.
}
}

\prf{
As in the previous proof, it suffices by Theorem \ref{t:abs_recov} to estimate the coherence $\Gamma(F,\Delta)$ for subsets $\Delta$ of the form $\Delta = \Delta_1 \cup \cdots \cup \Delta_D$, where $\Delta_d \subseteq I_d$ and $s_d = | \Delta_d | \leq \lambda s / D$ for $d=1,\ldots,D$.  Let $z \in \bbC^{N}$, $\| z \|_{\infty}=1$ and $i \in \{1,\ldots,N\}$.  If $B \sim F$ then
\eas{
\left | e^*_i B B^* P_{\Delta} z \right | &= \left | \sum^{D}_{d=1} \sum_{j \in \Delta_d} z_j \sum^{C}_{c=1} e^*_i H^*_c a a^* H_c e_j \right |
\\
& \leq \sum^{D}_{d=1} s_d \max_{j \in I_d} \left | \sum^{C}_{c=1} e^*_i H^*_c a a^* H_c e_j \right |
\\
& \leq \frac{\lambda s}{D} \| a \|^2_{\infty} \sum^{C}_{d=1} \max_{j \in I_d} \left | \sum^{C}_{c=1} \overline{h_{c,i}} h_{c,j} \right |.
}
Therefore $\Gamma_{1}(F,\Delta) \leq  \lambda \cdot s \cdot \mu \cdot \Upsilon_{\mathrm{identical}}$.  Similarly, 
\eas{
\bbE | e^*_i B B^* P_{\Delta} |^2 &= \bbE \left | e^*_i a \sum^{C}_{c=1} \overline{h_{c,i}} a^* H_c P_{\Delta} z \right |^2
\\
& \leq \| a \|^2_{\infty} \bbE \left | a^* \left ( \sum^{C}_{c=1} \overline{h_{c,i}} H_c P_{\Delta} z \right ) \right |^2
\\
& = \| a \|^2_{\infty} \nm{\sum^{C}_{c=1} \overline{h_{c,i}} H_c P_{\Delta} z }^2,
}
where in the final step we use the fact that $G$ is isotropic.  Hence
\eas{
\bbE | e^*_i B B^* P_{\Delta} |^2 & \leq \| a \|^2_{\infty} \sum^{D}_{d=1} \sum_{j \in \Delta_d} | z_j |^2 \left | \sum^{C}_{c=1} \overline{h_{c,i}} h_{c,j} \right |^2
\\
& \leq \| a \|^2_{\infty} \frac{\lambda s}{D} \sum^{D}_{d=1} \max_{j \in I_d} \left | \sum^{C}_{c=1} \overline{h_{c,i}} h_{c,j} \right |^2 
}
Since the $H_c$ are diagonal, the normalization condition $\sum^{C}_{c=1} H^*_c H_c = I$ is equivalent to $\sum^{C}_{c=1} | h_{c,i} |^2 = 1$, $i=1,\ldots,N$.  In particular,
\bes{
\left | \sum^{C}_{c=1} \overline{h_{c,i}} h_{c,j} \right | \leq \sqrt{\sum^{C}_{c=1} | h_{c,i} |^2 } \sqrt{  \sum^{C}_{c=1} | h_{c,j} |^2} = 1.
}
It now follows that $\bbE | e^*_i B B^* P_{\Delta} |^2 \leq \| a \|^2_{\infty} \lambda s \Upsilon_{\mathrm{identical}}$ and therefore $\Gamma_{2}(F,\Delta) \leq \lambda \cdot s  \cdot \mu \cdot \Upsilon_{\mathrm{identical}}$.  Combining this with the result for $\Gamma_{1}(F,\Delta)$ completes the proof.
}

This bound improves on our previous result \cite[Cor.\ 4.2]{Chun&Adcock:16arXiv-CS&PA}, since it depends on the quantity $\Upsilon_{\mathrm{identical}}$ whereas the bound Corollary 4.2 of \cite{Chun&Adcock:16arXiv-CS&PA} depends linearly on $C$.

\subsection{Discussion}
For distinct and identical sampling respectively, Corollaries \ref{c:distinct_bound} and \ref{c:identical_bound} provide optimal recovery guarantees, provided the partition $\cI$ and sensors profiles $H_c$ are such that $\Upsilon_{\mathrm{distinct}}$ and $\Upsilon_{\mathrm{identical}}$ are independent of $C$.  Note that $\Upsilon_{\mathrm{distinct}}, \Upsilon_{\mathrm{identical}} \leq C$ in general, which agrees with the worst-case bounds derived in \cite{Chun&Adcock:16arXiv-CS&PA}.  Yet, it is possible to construct large families of sensor profile matrices for which $\Upsilon_{\mathrm{distinct}}$ and $\Upsilon_{\mathrm{identical}}$ are independent of $C$, thus yielding optimal recovery.  We consider several such examples in \S \ref{sec:egs}.

Interestingly, $\Upsilon_{\mathrm{distinct}}$ and $\Upsilon_{\mathrm{identical}}$ are computable in $\ord{C N}$ and $\ord{C N^2}$ operations respectively.  Hence, optimal recovery can be easily checked numerically.  Thus, Corollaries \ref{c:distinct_bound} and \ref{c:identical_bound} give a practical means to approach the question of optimal design of sensor profiles, within the constraints of a particular application.

\section{Examples of diagonal sensor profiles}
\label{sec:egs}

We now introduce several different families of diagonal sensor profiles that lead to optimal recovery guarantees for both distinct and identical sampling.

\subsection{Piecewise constant sensor profiles}
\label{sec:eg:piecewise}
The following example was first presented in \cite{Chun&Adcock:16arXiv-CS&PA}.  Let $\cI = \{ I_1,\ldots,I_D\}$ be a partition of $\{1,\ldots,N\}$, where $D \leq C$, and suppose that $V = \{ V_{c,d}: c=1,\ldots,C, d = 1,\ldots,D \} \in \bbC^{C \times D}$ is an isometry, i.e.\ $V^* V = I$. 
Define the sensor profile matrices
\bes{
\label{Hc_pcwse_const_identical}
H_c = \sqrt{\frac{C}{M}} \sum^{D}_{d=1} V_{c,d} P_{I_d},
}
where, as in \S \ref{sec:general_setup}, $M=1$ for distinct sampling and $M=C$ for identical sampling.  Observe that $\sum^{C}_{c=1} H^*_c H_c = C/M\sum^{D}_{d=1} P_{I_d} \sum^{C}_{c=1} | V_{c,d} |^2 = C/M \sum^{D}_{d=1} P_{I_d} = (C/M) I$, so the profiles satisfy the respective joint isometry conditions.  Furthermore, in the distinct case
\bes{
\Upsilon_{\mathrm{distinct}} = \frac{C}{D} \max_{c=1,\ldots,C} \sum^{D}_{d=1} \max_{e=1,\ldots,D} | V_{c,e} | |V_{c,d} | \leq C \mu(V),
}
where $\mu(V) = \max_{c,d} |V_{c,d} |^2$ is the coherence of the matrix $V$.  Hence, for distinct sampling, we obtain an optimal recovery guarantee whenever $V$ is incoherent, i.e.\ $\mu(V) \lesssim C^{-1}$.\footnote{Since $V \in \bbC^{C \times D}$ is an isometry and $D \leq C$, its coherence satisfies $C^{-1} \leq \mu(V) \leq 1$.}  Note that this holds independently of the number of partitions $D$.  In particular, when $D=1$ we get optimal recovery of all $s$-sparse vectors.

Conversely, in the identical case
\eas{
\Upsilon_{\mathrm{identical}} &= \frac{C}{D} \max_{e=1,\ldots,D} \sum^{D}_{d=1} \left | \sum^{C}_{c=1} \overline{V_{c,e}} V_{c,d} \right |
\\
&= \frac{C}{D} \max_{e=1,\ldots,D} \sum^{D}_{d=1} \delta_{d,e} = \frac{C}{D}.
}
Hence we obtain an optimal recovery guarantee whenever the number of partitions $D$ is such that $C/D \lesssim 1$.  Note that this does not require $V$ to be incoherent, as in the case of distinct sampling.  However, it only ensures recovery of vectors that are sparse and distributed, as opposed to all sparse vectors.

\subsection{Banded sensor profile} 
\label{sec:eg:banded}
Let $\cI = (I_1,\ldots,I_D)$ be a partition and suppose that the $h_c$ are banded, i.e.\
\bes{
\supp(h_c) \subseteq \bigcup^{r_2}_{d=-r_1} I_{c+d},
}
for some fixed $r_1 \in \bbN$ and $r_2 \in \bbN$ (note that $I_{c+d} = 0$ if $c+d < 0$ or $c+d > C$).  Since $\sum^{C}_{c=1}\left | \overline{h_{c,i}}  h_{c,j} \right | \leq C/M$, where $M$ is as in the previous example, it follows that
\bes{
\Upsilon_{\mathrm{distinct}} , \Upsilon_{\mathrm{identical}} \leq C/D ( r_1+r_2+1). 
}
Hence in both cases we get an optimal recovery guarantee whenever $D$ is such that $C/D \lesssim 1$ and the bandwidth $r_1+r_2+1$ is independent of $C$.

\begin{figure}[t!]
\centering
\begin{tabular}{ccc}
\includegraphics[scale=0.55, trim=0.2em 0.4em 2.6em 2.8em, clip]{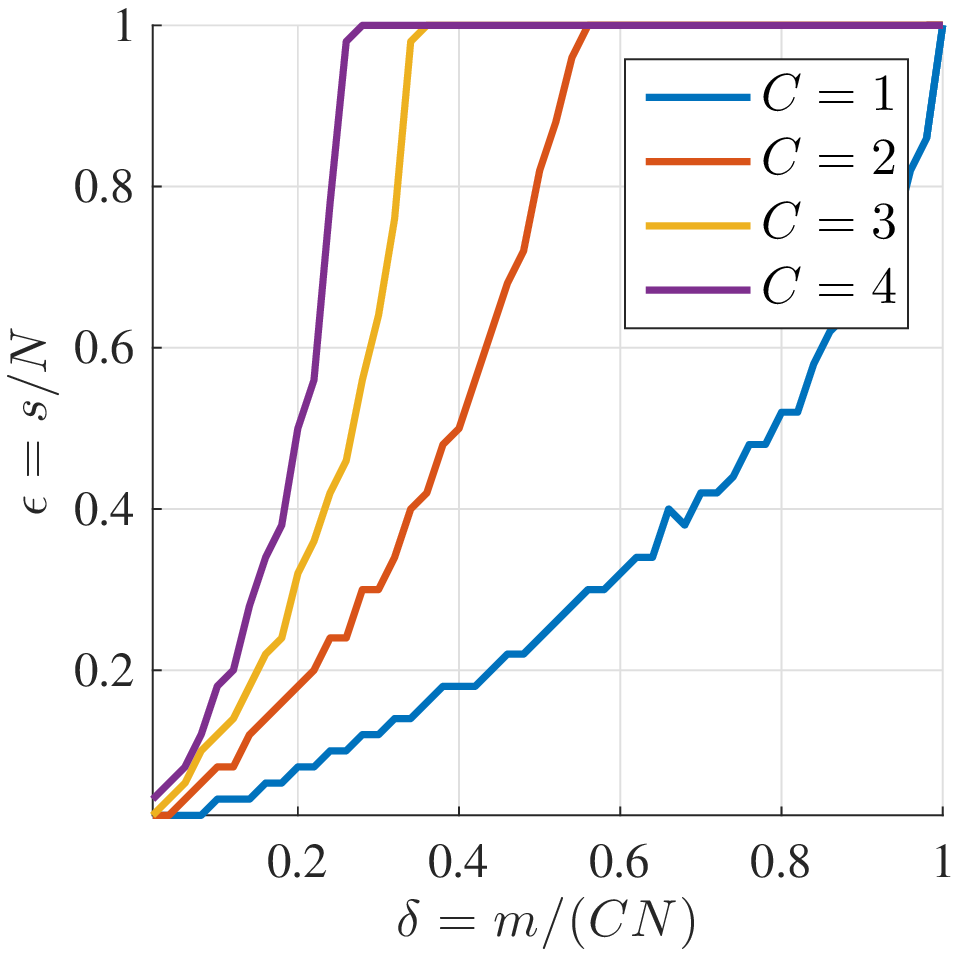} &
{} &
\includegraphics[scale=0.55, trim=0.2em 0.4em 2.6em 2.8em, clip]{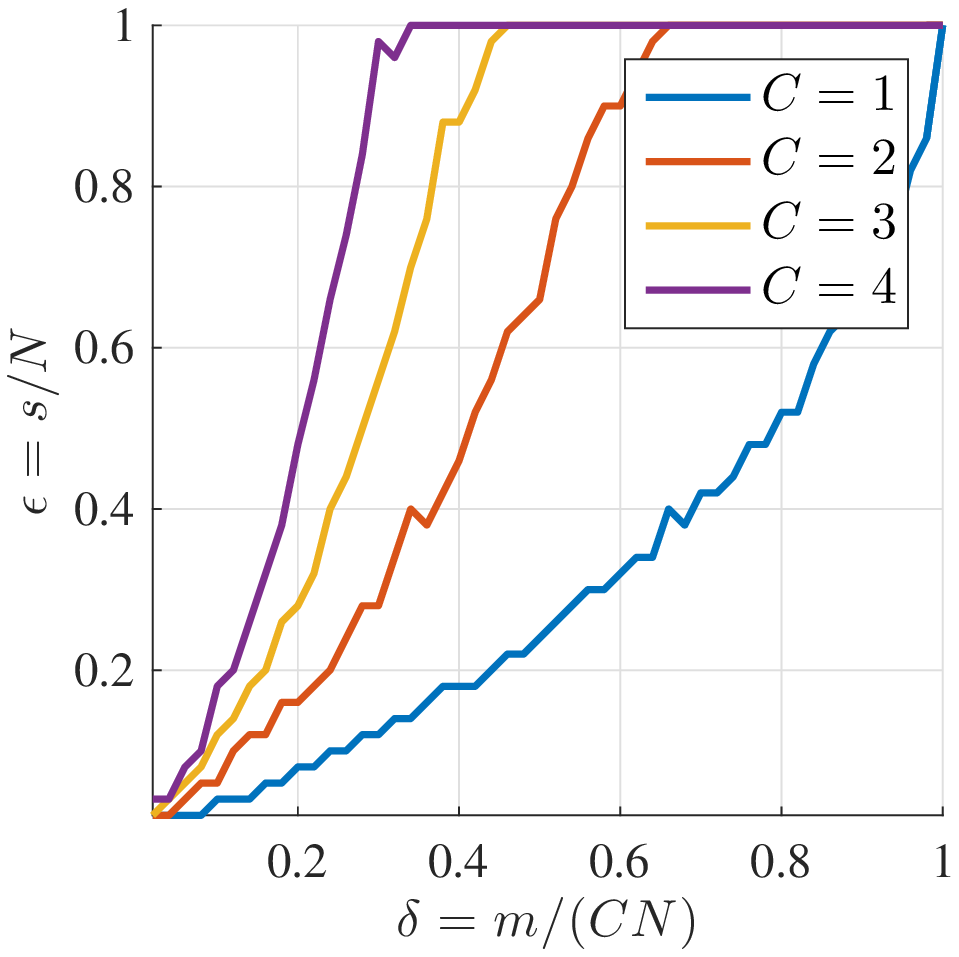} \\
{\small \qquad (a) Distinct sampling} & {\small} &{\small \qquad (b) Identical sampling} \\
\end{tabular}

\caption{
Empirical phase transitions for random Fourier sensing with banded diagonal sensor profile matrices and $C=1,2,3,4$ sensors. 
Phase transition curves with the empirical success probability $\approx 50 \%$ are presented (for details of phase transition experiment, see \cite{Chun&Adcock:16arXiv-CS&PA}).
For both sampling scenarios, the empirical probability of successful recovery increases as $C$ increases. The results are in agreement with our theoretical results. 
}
\label{fig:PTcurveFit_F_DiagH}
\end{figure}

A specific example of banded sensor profile stemming from applications is a smooth sensor profile with compact support \cite[Fig.\ 1(c)]{Chun&Adcock:16arXiv-CS&PA}. This corresponds to a sharply decaying coil sensitivity in a one-dimensional example of pMRI application;  see \cite{Chun&Adcock&Talavage:15TMI} for further details on the pMRI application.
For these sensor profiles, we set $D=C-1$ and 
\bes{
I_c = \{ (c-1) N / D + 1,\ldots, c N / D \},\qquad c=1,\ldots,D.
}
Note that this specific example corresponds to a banded sensor profile with $r_1 = 1$, $r_2 = 0$; therefore, $\Upsilon_{\mathrm{distinct}},\Upsilon_{\mathrm{identical}} \leq 2$ for any $C$, which leads to an optimal recovery guarantee.  This theoretical result is verified in Fig. \ref{fig:PTcurveFit_F_DiagH}(b), where empirical phase transition curves are computed for both types of sampling.

\section*{Acknowledgment}
BA wishes to acknowledge the support of Alfred P. Sloan Research Foundation and the Natural Sciences and Engineering Research Council of Canada through grant 611675.  BA and IYC acknowledge the support of the National Science Foundation through DMS grant 1318894.

\bibliographystyle{IEEEtran}
\bibliography{IEEEabrv,referenceBibs_Bobby}

\begin{thebibliography}{10}
\providecommand{\url}[1]{#1}
\csname url@samestyle\endcsname
\providecommand{\newblock}{\relax}
\providecommand{\bibinfo}[2]{#2}
\providecommand{\BIBentrySTDinterwordspacing}{\spaceskip=0pt\relax}
\providecommand{\BIBentryALTinterwordstretchfactor}{4}
\providecommand{\BIBentryALTinterwordspacing}{\spaceskip=\fontdimen2\font plus
\BIBentryALTinterwordstretchfactor\fontdimen3\font minus
  \fontdimen4\font\relax}
\providecommand{\BIBforeignlanguage}[2]{{%
\expandafter\ifx\csname l@#1\endcsname\relax
\typeout{** WARNING: IEEEtran.bst: No hyphenation pattern has been}%
\typeout{** loaded for the language `#1'. Using the pattern for}%
\typeout{** the default language instead.}%
\else
\language=\csname l@#1\endcsname
\fi
#2}}
\providecommand{\BIBdecl}{\relax}
\BIBdecl

\bibitem{Chun&Adcock:16arXiv-CS&PA}
\BIBentryALTinterwordspacing
I.~Y. Chun and B.~Adcock, ``Compressed sensing and parallel acquisition,''
  \emph{submitted to IEEE Trans. Inf. Theory}, Jan. 2016. [Online]. Available:
  \url{http://arxiv.org/abs/1601.06214}
\BIBentrySTDinterwordspacing

\bibitem{Chun&Adcock&Talavage:15TMI}
I.~Y. Chun, B.~Adcock, and T.~M. Talavage, ``Efficient compressed sensing
  {SENSE} {pMRI} reconstruction with joint sparsity promotion,'' \emph{IEEE
  Trans. Med. Imag.}, vol.~35, no.~1, pp. 354--368, Jan. 2016.

\bibitem{Chun&Adcock&Talavage:14EMBS_pMRI}
I.~Y. Chun, B.~Adcock, and T.~Talavage, ``Efficient compressed sensing {SENSE}
  parallel {MRI} reconstruction with joint sparsity promotion and mutual
  incoherence enhancement,'' in \emph{Proc. $36^{\text{th}}$ IEEE EMBS},
  Chicago, IL, Aug. 2014, pp. 2424--2427.

\bibitem{Pruessmann&Weiger&Scheidegger&Scheidegger:99MRM}
K.~P. Pruessmann, M.~Weiger, M.~B. Scheidegger, and P.~Boesiger, ``{SENSE}:
  sensitivity encoding for fast {$\textmd{MRI}$},'' \emph{Magn. Reson. Med.},
  vol.~42, no.~5, pp. 952--962, Jul. 1999.

\bibitem{Knoll&Clason&Bredies&Uecker&Stollberger:12MRM}
F.~Knoll, C.~Clason, K.~Bredies, M.~Uecker, and R.~Stollberger, ``Parallel
  imaging with nonlinear reconstruction using variational penalties,''
  \emph{Magn. Reson. Med.}, vol.~67, no.~1, pp. 34--41, Jan. 2012.

\bibitem{She&Chen&Liang&DiBella&Ying:14MRM}
H.~She, R.~R. Chen, D.~Liang, E.~V. DiBella, and L.~Ying, ``Sparse {BLIP}:
  {BL}ind {I}terative {P}arallel imaging reconstruction using compressed
  sensing,'' \emph{Magn. Reson. Med.}, vol.~71, no.~2, pp. 645--660, Feb. 2014.

\bibitem{Park&Wakin:12EJASP}
J.~Y. Park and M.~B. Wakin, ``A geometric approach to multi-view compressive
  imaging,'' \emph{EURASIP J. Adv. Signal Process.}, vol. 2012, no.~1, pp.
  1--15, Dec. 2012.

\bibitem{Traonmilin&etal:15JMIV}
Y.~Traonmilin, S.~Ladjal, and A.~Almansa, ``Robust multi-image processing with
  optimal sparse regularization,'' \emph{J. Math. Imaging Vis.}, vol.~51,
  no.~3, pp. 413--429, Mar. 2015.

\bibitem{Papoulis:77TCS}
A.~Papoulis, ``Generalized sampling expansion,'' \emph{IEEE Trans. Circuits
  Syst.}, vol.~24, no.~11, pp. 652--654, Nov. 1977.

\bibitem{Unser:PIEEE}
M.~Unser, ``Sampling-50 years after {S}hannon,'' \emph{Proc. IEEE}, vol.~88,
  no.~4, pp. 569--587, Apr. 2000.

\bibitem{Candes&Plan:11IT}
E.~J. Candes and Y.~Plan, ``A probabilistic and {RIP}less theory of compressed
  sensing,'' \emph{IEEE Trans. Inf. Theory}, vol.~57, no.~11, pp. 7235--7254,
  Nov. 2011.

\bibitem{Adcock&Hansen&Poon&Pi&Roman:13arXiv}
B.~Adcock, A.~C. Hansen, C.~Poon, and B.~Roman, ``Breaking the coherence
  barrier: a new theory for compressed sensing,'' \emph{arXiv pre-print
  cs.IT/1302.0561}, Feb. 2013.

\bibitem{Adcock&Hansen&Roman:14arXiv1403.6541}
B.~Adcock, A.~C. Hansen, and B.~Roman, ``A note on compressed sensing of
  structured sparse wavelet coefficients from subsampled {F}ourier
  measurements,'' \emph{arXiv pre-print math.FA/1403.6541}, Mar. 2014.

\end{thebibliography}

\end{document}